\documentclass[aps,superscriptaddress,showpacs,showkeys]{revtex4}
\usepackage[dvipdf]{graphicx}
\usepackage{amssymb}
\usepackage{amsbsy}
\usepackage{latexsym}

\newcommand{\be}{\begin{equation}}
\newcommand{\ee}{\end{equation}}
\newcommand{\ba}{\begin{eqnarray}}
\newcommand{\ea}{\end{eqnarray}}

\begin{document}

\title{Mixmaster  universe in the $z=3$
deformed Ho\v{r}ava-Lifshitz gravity}

\author{Yun Soo Myung}

\email{ysmyung@inje.ac.kr}

\affiliation {Institute of Basic Science and School of Computer
Aided Science, Inje University, Gimhae 621-749, Korea}

\author{Yong-Wan Kim}

\email{ywkim65@gmail.com}

\affiliation {Institute of Basic Science and School of Computer
Aided Science, Inje University, Gimhae 621-749, Korea}

\author{Woo-Sik Son}

\email{dawnmail@sogang.ac.kr}

\affiliation {Department of Physics and WCU-SSME Program Division, Sogang
University, Seoul 121-742, Korea}

\author{Young-Jai Park}

\email{yjpark@sogang.ac.kr}

\affiliation {Department of Physics and WCU-SSME Program Division, Sogang
University, Seoul 121-742, Korea}

\begin{abstract}
The $z=3$ deformed Ho\v{r}ava-Lifshitz gravity with coupling
constants $\omega$ and $\epsilon$   leads to a nonrelativistic
``mixmaster" cosmological model. The potential of  theory is given
by the sum of IR and UV potentials in the ADM Hamiltonian formalism.
It turns out that the presence of  the UV-potential cannot  suppress
chaotic behaviors existing in the IR-potential, which comes from
curvature anisotropy.
\end{abstract}



\maketitle

\section{Introduction}

Recently, quantum gravity at a Lifshitz point, which is
power-counting renormalizable and hence potentially UV complete, was
proposed by Ho\v{r}ava~\cite{ho1,ho2,ho3}. This theory for quantum
gravity is not intended to be a unified theory like string theory.

A hot issue of the Ho\v{r}ava-Lifshitz gravity is to answer to the
question of whether it can
 accommodate the Ho\v{r}ava scalar $\psi$,
  in addition to two degrees of freedom (DOF) for a massless graviton.
This additional scalar degree of freedom inevitably appears as a
result of the reduced symmetry of diffeomorphism known as
``foliation diffeomorphsim"~\cite{CNPS,LP,SVW,BPS1,KA,HKG}. The
authors~\cite{CNPS} have shown that without the projectability
condition, a perturbative general relativity  cannot be reproduced
in the IR-limit of the $z=3$ deformed Ho\v{r}ava-Lifshitz gravity
because of the strong coupling problem.
 With the projectability condition, the authors~\cite{SVW} have argued that $\psi$ is
propagating around the Minkowski space but it has a negative kinetic
term, showing a ghost mode.   Moreover, it was found that the
Ho\v{r}ava scalar is a ghost if the sound speed squared is
 positive (strong coupling problem)~\cite{BPS1}.  Even the Lorentz-violating mass term was
 included,  the mass term did not
cure  the ghost problem of the Ho\v{r}ava scalar~\cite{M-massive}.
In order to resolve the strong coupling
 problem, Blas, Pujolas, and Sibiryakov have proposed an extended
 version of the Ho\v{r}ava-Lifshitz gravity where the lapse function
 $N$ may depend on the spatial coordinate $r$ (the theory is not
 projectable) and thus, terms of $\partial_i \ln N$ are included in
 the action~\cite{BPS2}. It was argued that this extended version
 are free from the strong coupling pathology. However, the extended
 theory could still suffer from the strong coupling at low
 energies in the kinetic term~\cite{PS}, but it can be evaded by including higher
 spatial derivative terms~\cite{BPS3}.
  Hence, up to now, the strong coupling issue is not completely
  resolved even though the extended theory was seriously considered.

Specific cosmological implications of the $z=3$
Ho\v{r}ava-Lifshitz gravity with the Friedmann-Robertson-Walker
(FRW) metric based on isotropy and homogeneity have recently been
shown in~\cite{cal,KK,muk}, including homogeneous vacuum solution
with chiral primordial gravitational waves~\cite{TS} and nonsingular
cosmological evolution with the big bang of standard and
inflationary universe replaced by a matter
bounce~\cite{Bra,Rama,LS}. As far as the isotropic solutions are
concerned, there is no difference between $z=2$~\cite{ho1} and
$z=3$~\cite{ho2} Ho\v{r}ava-Lifshitz gravities because the Cotton
tensor vanishes when using the isotropic  FRW metric. Furthermore,
one has introduced the  $z=3$  deformed Ho\v{r}ava-Lifshitz gravity
to find asymptotically flat background~\cite{KS,Myungbh}.

On the other hand, the equations of general relativity lead to
singularities when we look at the equations backwards the origin of
time. Especially, we concentrate on a temporal singularity of the
solutions to the Einstein equations for the mixmaster model (Bianchi
IX Universe) describing an anisotropic and homogeneous cosmology. It
was well known that the approach to singularity shows a chaotic
behavior. The mixmaster
universe~\cite{mix1,mix2,mix3,mix4,cl,mix5,mix6,mix7} could be
described by a Hamiltonian dynamical system in a 6D phase space.
Belinsky, Khalatnikov, and Lifshitz (BKL) had conjectured that this
6D phase system could be well approximated by a 1D discrete Gauss
map that is known to be chaotic as one approaches the
singularity~\cite{BKL}. Chernoff and Barrow have suggested that the
mixmaster 6D phase space could be split into the product of a 4D
phase space and a 2D phase space having regular
variables~\cite{mix3}. Following Cornish and Levin~\cite{cl}, Lehner
and Di Menza have  found that the chaos in the mixmaster universe is
obtained for the Hamiltonian system with potential having fixed
walls, which describes the curvature anisotropy~\cite{mix6}.

However, it turned out  that the mixmaster chaos could be suppressed
by (loop) quantum effects~\cite{BD,Bo}.
In the loop quantum cosmology, the effective potential at decreasing
volume labeled by ``discreteness $j$"  are significantly changed in
the vicinity of (0,0)-isotropy point in the anisotropy plane
$(\beta_+,p_+)$. The potential at larger volumes exhibits a
potential wall of finite height and finite extension. As the volume
is decreased, the wall moves inward and its height decreases.
Progressively, the wall disappears completely making the potential
negative everywhere at a dimensionless volume of $(2.172j)^{3/2}$ in
the Planck units. Eventually, the potential approaches zero from
below. This shows that classical reflections will stop after a
finite amount of time, implying that classical arguments about chaos
are inapplicable. Once quantum effects are taken into account, the
reflections stop just when the volume of a given patch is about the
size of Planck volume.

We point out that loop quantum gravity is a non-perturbative and background independent
canonical quantization of general relativity, while loop quantum
cosmology is a cosmological mini-superspace model quantized with
methods of loop quantum gravity. Hence the discreteness of spatial
geometry and the simplicity of setting allow for complete study of
cosmological evolution. The difference between loop quantum
cosmology and other approaches of quantum cosmology is that the
input is plugged by a full quantum gravity theory, which introduces a
discreteness to space-time. That is, in order to quantize general
relativity, this discreteness manifests itself as quanta of space.

Recently, we have investigated the $z=2$ deformed
Ho\v{r}ava-Lifshitz gravity with coupling constant $\omega$ which
leads to a nonrelativistic ``mixmaster" cosmological
model~\cite{MKSP}. We have obtained that for $\omega>0$, there
always exists chaotic behavior. This contrasts  to the case of the
loop mixmaster dynamics based on loop quantum cosmology~\cite{Bo},
where the mixmaster chaos is suppressed by loop quantum
effects~\cite{BD}. We recognize that  the role of UV coupling
parameter $\omega$ is intrinsically different from the area quantum
number $j$ of the loop quantum cosmology which controls the volume
of the universe.  In our case, time variable (related to the volume
of $V=e^{3\alpha}$) as well as two physical degrees of anisotropy
$\beta_{\pm}$ are treated in the standard way without quantization.
However, in the loop quantum framework, all three scale factors were
quantized using the loop techniques. Hence two are quite different:
the  potential wells at the origin never disappear for any
$\omega>0$ in the $z=2$ Ho\v{r}ava-Lifshitz gravity, while in the
loop quantum gravity the height of potential wall rapidly decreases
until they disappears completely as the Planck scale is reached.

On the other hand, it was interestingly shown that adding 4D
curvature squared term $(^{4}R)^2$ (and possibly other) curvature
squared terms to the Einstein gravity leads to an interesting
result that the chaotic behavior is absent~\cite{BC1,BC2,BC3}.
Hence it is very curious to see why $(^{4}R)^2$ does suppress
chaotic behavior but 3D curvature squared terms of
$\frac{3}{4\omega}R^2-\frac{2}{\omega}R_{ij}R^{ij}$ does not
suppress chaotic behavior. It was argued that the absence of chaos
in  covariant higher curvature generalization of Einstein gravity
$f(^{4}R)$ is due to the presence of a scalar $\varphi=\log
f'(^{4}R)$~\cite{BBLP}.  This scalar  slows down the velocity of
the point particle (the universe) relative to the moving walls and
thus, the universe will bounce back only if it moves not too
oblique relative to the walls. A few of collisions are sufficient
to make it so oblique that it will not bounce off another wall.
The universe will enter quickly in a definite Kasner trajectory
and stay there all the time in its approach to the singularity.
Hence, the evolution of the universe is not chaotic.

Hence it is very interesting to investigate cosmological application
of the $z=3$ Ho\v{r}ava-Lifshitz gravity in conjunction with the
mixmaster universe based on the anisotropy and homogeneity because
this Ho\v{r}ava-Lifshitz gravity may be regarded  as a strong
candidate for quantum gravity. The mixmaster universe in the $z=3$
Ho\v{r}ava-Lifshitz gravity was discussed in Ref.\cite{BBLP}.
However, the authors~\cite{BBLP} have focused on the Cotton bilinear
term $C_{ij}C^{ij}$ only and thus, have briefly  sketched possible
dynamical behaviors of the universe when approaching the initial
singularity.

For the isotropic case of the $z=3$  Ho\v{r}ava-Lifshitz gravity,
the $k=1$ FRW universe with dark radiation and dust matter ($w=0$)
has led to a matter bounce. If the Ho\v{r}ava-Lifshitz gravity is
true, the universe did not bang-it bounced. That is, a universe
filled with matter will contract down to a small but finite size and
then bonce again, giving us the expanding universe  that we see
today. This bounce scenario indicates a key feature of the
Ho\v{r}ava-Lifshitz gravity, showing an essential difference from
the big bang scenario. On the other hand, for the anisotropic case
of Einstein gravity, the mixmaster universe filled with stiff matter
($w=1$) has led to a non-chaotic universe because there is a slowing
down of particle velocity, which is unable to reach any more the
walls after some time in the moving wall picture.  Hence, an urgent
issue for an anisotropic mixmaster universe  is to see whether
there exists a mechanism to slow down the particle velocity  in the
$z=3$ Ho\v{r}ava-Lifshitz gravity.

In this work, we wish to find whether a mechanism to stop  chaotic
behaviors exists in  the Ho\v{r}ava-Lifshitz gravity. We will analyse
the $z=3$ deformed Ho\v{r}ava-Lifshitz gravity
without cosmological constant to make the situation simple.

\section{$z=3$ deformed Ho\v{r}ava-Lifshitz gravity}
In order to get an associated Hamiltonian within the ADM
formalism~\cite{adm} of the  $z=3$ deformed Ho\v{r}ava-Lifshitz
gravity~\cite{ho1,KS,Myungch}, we have to find three potentials in
6D phase space: IR-potential $V_{IR}$ from 3D curvature $R$ and two
UV-potentials: $V^{(I)}_{UV}$ from curvature squared terms of $R^2$
and $R_{ij}R^{ij}$ with UV coupling parameter  $\omega$ and
$V^{(II)}_{IR}$ from $C_{ij}R^{ij}$ and $C_{ij}C^{ij}$ with
additional coupling constant $\epsilon$.

We start with the action of the  $z=3$  deformed Ho\v{r}ava-Lifshitz
gravity~\cite{ho1,KS}
\begin{eqnarray}
 S_\lambda = \int dtd^3x\sqrt{g}N\left[\frac{2}{\kappa^2}(K_{ij}K^{ij}-\lambda K^2)
          + \mu^4 R
          + \frac{\kappa^2\mu^2(1-4\lambda)}{32(1-3\lambda)}R^2
           -\frac{\kappa^2\mu^2}{8}R_{ij}R^{ij}
           +\frac{\kappa^2\mu}{2\eta^2}C_{ij}R^{ij}
           -\frac{\kappa^2}{2\eta^4}C_{ij}C^{ij}\right]
\end{eqnarray}
with four parameters $\kappa,~\mu,~\lambda$, and $\eta$. In the case
of $\lambda=1$, the above action leads to
\begin{eqnarray}
\label{action}
 S_{\lambda=1}= \int  dtd^3x \sqrt{g}N \mu^4 \Bigg[\frac{1}{c^2}(K_{ij}K^{ij}-K^2)
              +R+\frac{3}{4\omega}R^2-\frac{2}{\omega}R_{ij}R^{ij}
              +\frac{8\sqrt{2}}{\omega^{7/6}\epsilon}C_{ij}R^{ij}
              -\frac{16}{\omega^{4/3}\epsilon^2}C_{ij}C^{ij}\Bigg]
\end{eqnarray}
where the two UV coupling parameters $\omega=16\mu^2/\kappa^2$ and
$\epsilon=\eta^2c^{1/3}$ are introduced to control curvature and
Cotten squared terms~\cite{Myungch}. In the limit of $\omega\to
\infty ~(\kappa^2 \to 0)$, $S_{\lambda=1}$ reduces to Einstein
gravity (GR) with the speed of light $c^2=\kappa^2\mu^4/2$ and
Newton's constant $G=\kappa^2/(32\pi c)$.

Now, let us introduce the metric for the mixmaster universe  to
distinguish between expansion (volume change: $\alpha$) and
anisotropy (shape change: $\beta_{ij}$) as follows
\begin{equation} \label{metric}
  ds^2=-dt^2+e^{2\alpha}e^{2\beta_{ij}}\sigma^i \otimes \sigma^j,
\end{equation}
where $\sigma^i$ are the 1 forms given by
\begin{eqnarray}
 && \sigma^1=\cos\psi d\theta+\sin\psi\sin\theta d\phi,\nonumber\\
 && \sigma^2=\sin\psi d\theta-\cos\psi\sin\theta d\phi,\nonumber\\
 && \sigma^3=d\psi+\cos\theta d\phi
\end{eqnarray}
on the three-sphere parameterized by Euler angles
($\psi,\theta,\phi$) with $0\leq\psi<4\pi$, $0\leq\theta<\pi$, and
$0\leq\phi<2\pi$. The shape change $\beta_{ij}$ is a $3\times 3$ traceless
symmetric tensor with det[$e^{2\beta_{ij}}]=1$ expressed in terms
of two independent shape parameters $\beta_\pm$ as
\begin{equation} \label{betapm}
 \beta_{11}=\beta_++\sqrt{3}\beta_-,~~\beta_{22}=\beta_+-\sqrt{3}\beta_-,~~\beta_{33}=-2\beta_+.
\end{equation}
Then, the evolution of the universe can be described by giving $\beta_\pm$ as
function of $\alpha$. Note that the $k=1$ FRW universe is the
special case of $\beta_\pm=0$.

Now we concentrate on the behavior near singularity. Then, the empty
space without matter  is sufficient to display the generic local
evolution close to  singularity  because the terms due to  dust
matter or radiation are negligible near singularity.

Before we proceed, let us consider the Einstein gravity.
 Using Eq.
(\ref{metric}), the 3D curvature takes the form
\begin{equation}
 R= -12e^{-2\alpha}V_{IR}(\beta_+,\beta_-),
\end{equation}
where the IR-potential of curvature anisotropy is given by
\begin{eqnarray}
  V_{IR}(\beta_+,\beta_-) = \frac{1}{24}\Big[2e^{4\beta_+}\cosh(4\sqrt{3}\beta_-)+e^{-8\beta_+}\Big]
                          - \frac{1}{12}\Big[2e^{-2\beta_+}\cosh(2\sqrt{3}\beta_-)+e^{4\beta_+}\Big].
\end{eqnarray}
\begin{figure}[t]
\includegraphics{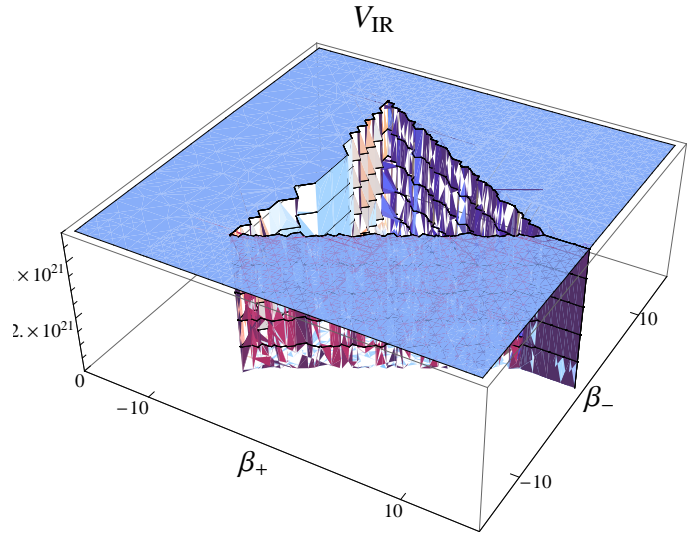}
\caption{The typical potential well $V_{IR}$ for fixed $\alpha=1$.
Three canyon lines are located at $\beta_-=0$ and $\beta_-=\pm
\sqrt{3}\beta_+$. } \label{fig1.eps}
\end{figure}
\begin{figure}[t]
\includegraphics{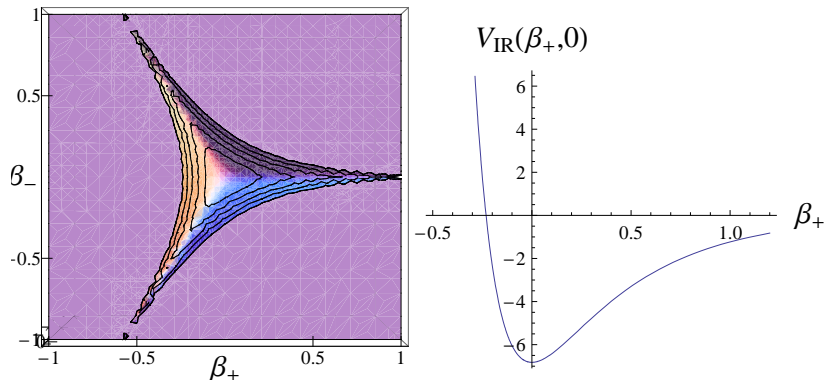}
\caption{The equipotential curves of $V_{IR}$. Left panel shows
equipotential curves viewed from the top and right panel indicates
the shape of potential $V_{IR}$, which has  no local maxima along
the canyon line $\beta_-=0$, compared to $V_{UV}^{(I)}$ and
$V_{UV}^{(II)}$. } \label{fig2.eps}
\end{figure}
Figure \ref{fig1.eps} depicts  a typical IR-potential  where three
canyon lines located at $\beta_-=0$ and $\beta_-=\pm \sqrt{3}
\beta_+$, showing an axial symmetry. It has the shape of an
equilateral triangle in the space labeled by ($\beta_+,\beta_-$) and
exponentially steep walls far away from the origin. As is shown in
Fig. \ref{fig2.eps}, the potential is the well   close to the origin
$(0,0)$: left panel shows equipotential curves viewed from the top,
while right panel is the shape of potential.
The origin (0,0), which corresponds to the isotropic case, is the
global minimum with negative value. Near the origin, the
IR-potential takes concentric forms of equipotential curves as
\begin{equation}
\label{zzvir}
V_{IR}(0,0)\approx -\frac{1}{8}+(\beta^2_++\beta^2_-).
\end{equation}
On the other hand, the asymptotic form of the IR-potential for  the
case of $\beta_- \ll 1 $   is either $V_{IR}\approx
2e^{4\beta_+}\beta^2_-$ if $\beta_+ \to \infty$  or $V_{IR}\approx
\frac{1}{24}e^{-8\beta_+}$ if $\beta_+\to -\infty$. That is, the
both walls grow exponentially. $V_{IR}$ has  no local maxima along
$\beta_-=0$, compared to $V_{UV}^{(I)}$ and $V_{UV}^{(II)}$. For
$\beta_-=0$, the IR-potential approaches zero from below if
$\beta_+\to \infty$. Hence the point particle with positive energy
$E>0$ can  escape to infinity along the canyon lines. The smallest
deviation from axial symmetry will turn the particle against the
infinitely steep walls.

The evolution of the universe is described by the motion of a point
$\beta=(\beta_+,\beta_-)$ as a function of $\alpha$ using the
time-dependent Lagrangian. The exponential wall picture of
the IR-potential implies that a particle (the universe) runs through
almost free (Kasner) epochs where the potential could be neglected,
and it is reflected  at the walls, resulting infinite number of
oscillations. This  implies that Einstein gravity  with the
IR-potential $V_{IR}$  shows  chaotic behaviors when the singularity
is approached~\cite{dhn}.

\begin{figure}[t]
\includegraphics{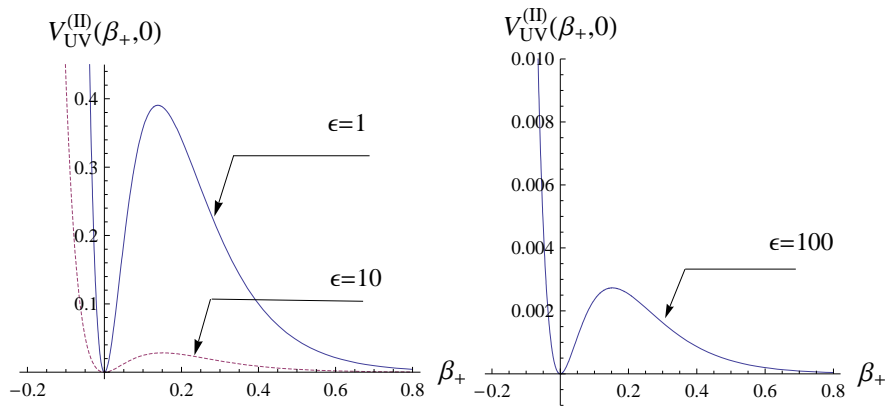}
\caption{The UV-potentials of $V^{(II)}_{UV}$ for
$\alpha=0,~\beta_-=0,$ and $\omega=1$ with $\epsilon=1$ (solid),
$10$ (dotted), and $100$ (dashed), respectively. There exist local
maxima $V_{lm}$, compared to the IR-potential $V_{IR}$.}
\label{fig3.eps}
\end{figure}

The action (\ref{action}) provides  the time-dependent Lagrangian
\begin{eqnarray}
\label{actionA}
  {\cal L}_{\lambda=1} = (4\pi)^2 \mu^4 e^{3\alpha}\left[-6(\dot{\alpha}^2-\dot{\beta}^2_+-\dot{\beta}^2_-)
           -12e^{-2\alpha}V_{IR}(\beta_+,\beta_-)
        - \frac{e^{-4\alpha}}{16\omega}V^{(I)}_{UV}(\beta_+,\beta_-)
        - e^{-6\alpha} V^{(II)}_{UV}(\beta_+,\beta_-) \right],
\end{eqnarray}
where the dot denotes $\frac{d}{cdt}$.  One needs to introduce an
emergent speed of light $c$ in order to see the UV behaviors, while
for the IR behaviors, one chooses $c=1$ simply. Here, the
UV-potential $V^{(I)}_{UV}$ is defined from  the curvature squared
terms  as
\begin{equation}
\frac{3}{4\omega}R^2-\frac{2}{\omega}R_{ij}R^{ij} \equiv
 -\frac{e^{-4\alpha}}{16\omega} V^{(I)}_{UV},
\end{equation}
where the UV-potential $V^{(I)}_{UV}$ takes the form of
\begin{eqnarray}
  V^{(I)}_{UV}(\beta_+,\beta_-) &\equiv& -\left[40 \Big(e^{8\beta_+}\cosh(4\sqrt{3}\beta_-)
                +e^{2\beta_+}\cosh(6\sqrt{3}\beta_-)+e^{-10\beta_+}\cosh(2\sqrt{3}\beta_-)\Big)
                -40e^{2\beta_+}\cosh(2\sqrt{3}\beta_-)\right.\nonumber\\
               && \left.+4e^{-4\beta_+}\cosh(4\sqrt{3}\beta_-)+2e^{8\beta_+}-20e^{-4\beta_+}
                   - 42e^{8\beta_+}\cosh(8\sqrt{3}\beta_-)
                   -21e^{-16\beta_+}\right].
\end{eqnarray}
On the other hand,  the other UV-potential $V^{(II)}_{UV}$ is found
from the Cotton terms as
\begin{equation}
\frac{8\sqrt{2}}{\omega^{7/6}\epsilon}C_{ij}R^{ij}
              -\frac{16}{\omega^{4/3}\epsilon^2}C_{ij}C^{ij}\equiv  -e^{-6\alpha} V^{(II)}_{UV}(\beta_+,\beta_-)
\end{equation}
with
\begin{eqnarray}
V^{(II)}_{UV}(\beta_+,\beta_-) &\equiv&
V^{CR}_{UV}(\beta_+,\beta_-)+V^{CC}_{UV}(\beta_+,\beta_-)
\nonumber \\
&=& \frac{8\sqrt{2}e^{\alpha}}{\omega^{7/6}\epsilon}
                \left[e^{-20\beta_+}+e^{-8\beta_+}-2e^{-14\beta_+}\cosh(2\sqrt{3}\beta_-)\right.\nonumber\\
                &&~~~~~~~~ +\left.2e^{4\beta_+}(\cosh(4\sqrt{3}\beta_-)-\cosh(8\sqrt{3}\beta_-))
                -2e^{10\beta_+}(\cosh(6\sqrt{3}\beta_-)-\cosh(10\sqrt{3}\beta_-))\right]
                \nonumber\\
                &-& \frac{8}{\omega^{4/3}\epsilon^2}
                \left[6-3e^{-24\beta_+}+6e^{-18\beta_+}\cosh(2\sqrt{3}\beta_-)
                -e^{-12\beta_+}(1+2\cosh(4\sqrt{3}\beta_-))\right.\nonumber\\
                &&~~~~~~~~ -
                4e^{-6\beta_+}(\cosh(2\sqrt{3}\beta_-)-\cosh(6\sqrt{3}\beta_-))
                -4\cosh(4\sqrt{3}\beta_-)-2\cosh(8\sqrt{3}\beta_-)\nonumber\\
                && ~~~~~~~~-2e^{6\beta_+}(2\cosh(2\sqrt{3}\beta_-)+\cosh(6\sqrt{3}\beta_-)-3\cosh(10\sqrt{3}\beta_-))\nonumber\\
                && ~~~~~~~~+2\left.
                e^{12\beta_+}(1-\cosh(4\sqrt{3}\beta_-)+3\cosh(8\sqrt{3}\beta_-)-3\cosh(12\sqrt{3}\beta_-))\right].
\end{eqnarray}
We have thoroughly studied the $V^{(II)}_{UV}=0$ case of the $z=2$
deformed Ho\v{r}ava-Lifshitz gravity in~\cite{MKSP}, indicating that
 chaotic behavior  persists,  as the Einstein gravity did show.  Thus, we point out that a key
feature of the $z=3$ deformed Ho\v{r}ava-Lifshitz gravity  is the
presence of the UV-potential $V^{(II)}_{UV}$. As was mentioned
in~\cite{BBLP}, the Cotton bilinear term $V^{CC}_{UV}$ contributes
to $V^{(II)}_{UV}$ without $\alpha$.  Near the origin
$(\beta_+,\beta_-)=(0,0)$, the UV-potential of $V^{(II)}_{UV}$ is
approximated  by
\begin{equation}
 V^{(II)}_{UV}(\beta_+,\beta_-) \approx
 \left(\frac{288\sqrt{2}e^\alpha}{\omega^{7/6}\epsilon}+\frac{864}{\omega^{4/3}\epsilon^2}\right)
 \left(\beta^2_++\beta^2_-\right).
\end{equation}
This means that  $V^{(II)}_{UV}(0,0)=0$ has no contribution to the
isotropic point (0,0), in contrast to  the IR-potential
$V_{IR}(0,0)=-1/8$ and UV-potential $V^{(I)}_{UV}(0,0)=-3$. Fig.
\ref{fig3.eps} indicates  shape changes  of the UV-potential
$V^{(II)}_{UV}$ for different values $\epsilon$,  showing different
local maxima $V_{lm}(\beta_+,0,1,\epsilon)$ for different
$\epsilon$. The asymptotic form for $\beta_-\ll 1$ is either
$V^{(II)}_{UV}\approx \frac{6144\beta^2_-
e^{12\beta_+}}{\omega^{4/3}\epsilon^2}$ if
$\beta_+\rightarrow\infty$, or $V^{(II)}_{UV}\approx
\frac{24e^{-24\beta_+}}{\omega^{4/3}\epsilon^2}$ if
$\beta_+\rightarrow-\infty$. An important point is that unlike
$V_{IR}$ and $V^{(I)}_{UV}$, the asymptotic form of
$V^{(II)}_{UV}\to V^{CC}_{UV}$ is independent of volume change
$\alpha$. As before, for $E>V_{lm}$, the point particle can escape
to infinity along the canyon lines $\beta_-=0$ and $\beta_-=\pm
\sqrt{3}\beta_+$.

In order to appreciate implications of chaotic approach to the
 $z=3$ deformed Ho\v{r}ava-Lifshitz gravity, we have to calculate the
Hamiltonian density by introducing three canonical momenta as
\begin{equation}
  p_\pm=\frac{\partial {\cal L}_{\lambda=1}}{\partial\dot{\beta}_\pm}=12(4\pi)^2\mu^4e^{3\alpha}\dot{\beta}_\pm,
  ~~~p_\alpha=\frac{\partial {\cal
  L}_{\lambda=1}}{\partial\dot{\alpha}}=-12(4\pi)^2\mu^4e^{3\alpha}\dot{\alpha}.
\end{equation}
The normalized canonical Hamiltonian in 6D phase space is given by
\begin{eqnarray}
  {\cal H}_{6D} &=&\frac{1}{2}(p^2_++p^2_--p^2_\alpha)
             +e^{4\alpha}\Big(V_{IR}+\frac{e^{-2\alpha}}{192 \omega}V^{(I)}_{UV}
             +\frac{e^{-4\alpha}}{12}V^{(II)}_{UV}\Big)\nonumber \\
  \label{potalpha}           &\equiv
             &\frac{1}{2}(p^2_++p^2_--p^2_\alpha)+V_{\alpha}(\beta_+,\beta_-,\omega,\epsilon),
\end{eqnarray}
where we have redefined ${\cal
H}_{6D}=12(4\pi)^2\mu^4e^{3\alpha}{\cal H}_c$ using the canonical
Hamiltonian ${\cal H}_c$, and chosen the parameter
$12(4\pi)^2\mu^4=1$ for simplicity.  Then, the Hamiltonian equations
of motion are
\begin{eqnarray}
\label{betapmeq} &&
 \dot{\beta}_\pm=p_\pm,
 ~~\dot{p}_\pm=-e^{4\alpha}\frac{\partial V_{IR}}{\partial\beta_\pm}
               -\frac{e^{2\alpha}}{192\omega}\frac{\partial V^{(I)}_{UV}}{\partial\beta_\pm}
               -\frac{1}{12}\frac{\partial V^{(II)}_{UV}}{\partial\beta_\pm},\\
\label{alphaeq} &&
 \dot{\alpha}=-p_\alpha,~~\dot{p}_\alpha
 =-4e^{4\alpha}V_{IR}
  -\frac{e^{2\alpha}}{96\omega}V^{(I)}_{UV}-\frac{1}{12}V^{CR}_{UV}
\end{eqnarray}
in 6D phase space.

\section{Isotropic evolution}
Now, let us see what happens in the isotropic point of (0,0) where the
$k=1$ FRW universe pops up. Since the isotropic potential does not receive
any contribution from the Cotton tensor, it is given by
\begin{equation}
V_{\alpha}(0,0,\omega)=-\Big(\frac{e^{4\alpha}}{8}+\frac{e^{2\alpha}}{64
\omega}\Big).
\end{equation}
From the Hamiltonian constraint of ${\cal H}_{6D}\approx0$, we have
the first Friedmann equation
\begin{equation}
\dot{\alpha}^2=-\frac{1}{4}\Big(\frac{1}{e^{2\alpha}}+\frac{1}{8\omega}\frac{1}{e^{4
\alpha}}\Big).
\end{equation}
Introducing the scaling factor $a=2e^{\alpha}$ with
$H=\frac{\dot{a}}{a}=\dot{\alpha}$, the above equation leads to
\begin{equation}
H^2=-\Big(\frac{1}{a^2}+\frac{1}{2\omega}\frac{1}{a^4}\Big),
\end{equation}
which is the same equation appeared for the Ho\v{r}ava-Lifshitz
cosmology~\cite{Bra,Rama,LS,Myungch}. The second term of right
handed side represents the dark radiation with negative energy
density. This means that the universe cannot evolve isotropically in
vacuum without turning on some shearing components. Adding a matter
density of $\rho=\frac{\rho_0}{a^{3(1+w)}}$ to the above equation
leads to
\begin{equation}
H^2=-\Big(\frac{1}{a^2}+\frac{1}{2\omega}\frac{1}{a^4}\Big)+\frac{\rho_0}{a^{3(1+w)}}.
\end{equation}
The solution to this equation can be obtained for $-1/3<\omega
<1/3$. Neglecting the first term of curvature, there can be a bounce
in $a$ that replaces the initial singularity of the universe. This
is the only case that dark radiation with negative energy density
can grow with respect to a regular matter energy density.

However, small derivations from isotropy will be dominant in the
small volume limit of $a\to 0(\alpha \to -\infty)$  because the
Cotton bilinear term, which is independent of $\alpha(a)$, kicks in
$V_\alpha(\beta_+,\beta_-,\omega,\epsilon)$ and washes away the
effects of dark radiation term. The kinetic energy of anisotropy
parameter $\beta_\pm$ also contributes to the universe evolution.
This implies that the cosmological bounce is unstable against
anisotropy and the universe can be in singular state of Kasner
universe. In the next section, we wish to study the mixmaster
universe of the $z=3$ Ho\v{r}ava-Lifshitz gravity explicitly.

\section{Chaotic behaviors in reduced 4D phase space}

 Chernoff and Barrow have showed that the mixmaster 6D phase
space could be split into the product of a
 4D phase space showing chaotic behavior and a 2D phase space showing regular
behavior~\cite{mix3}. Hence, we confine  the dynamical system to a
4D phase space describing the 4D static billiard in this section.
Setting $\alpha = 1$, let us consider the motion of a particle (the universe) of coordinates
($\beta_+,\beta_-$) under the potential of
\begin{equation}
 V(\beta_+,\beta_-,\omega,\epsilon)
 =e^4\Bigg[V_{IR}+\frac{V^{(I)}_{UV}}{192 e^2 \omega}+\frac{V^{(II)}_{UV}}{12 e^4}\Bigg].
\end{equation}
This potential has the symmetry of an equilateral  triangle
reflecting the equivalence of three axes in the metric
(\ref{betapm})~\cite{mix1}. Explicitly, a particle is moving in the
potential with exponential walls bounding a triangle. We mention
again that near the origin (0,0), the potential
$V(\beta_+,\beta_-,\omega,\epsilon)$ takes approximately the form of
\begin{equation}  \label{zzv} V(0,0,\omega,\epsilon) \approx
-\Big(\frac{e^{4}}{8}+\frac{e^2}{64\omega}\Big)+
\Big(e^4+\frac{17e^2}{4\omega}+\frac{24\sqrt{2}e}{\omega^{7/6}\epsilon}
+\frac{72}{\omega^{4/3}\epsilon^2}\Big)\Big(\beta^2_++\beta^2_-\Big).\end{equation}
Comparing (\ref{zzv}) with (\ref{zzvir}), the former reduces to the
latter up to $e^4$ in the IR-limit of $ \omega \to \infty$. It
turned out that adding the UV-potential $V^{(I)}_{UV}$  makes the
potential well deeper, compared to $V^{(II)}_{UV}$.

An important thing is to check whether the inflection point at the
origin of $(\beta_+,\beta_-)=(0,0)$ appears as $\omega$ varies,
which might show a signal of  changing  from chaotic to non-chaotic
behavior.  This inflection point is determined by the condition of
\begin{equation}
V''(\beta_+,0,\omega,\epsilon)|_{\beta_+=0}=V''(0,\beta_-,\omega,\epsilon)|_{\beta_-=0}=0,
\end{equation}
which leads to an algebraic equation
\begin{equation} \label{inflection}
2e^4+\frac{17e^2}{2\omega}+\frac{48\sqrt{2}e}{\omega^{7/6}\epsilon}+\frac{144}{\omega^{4/3}\epsilon^2}=0.
\end{equation}
However, we find that there is no such  positive solution $\omega$
and $\epsilon$ to Eq. (\ref{inflection}).  This shows clearly that
an inflection point could not be developed  by adjusting $\omega$
and $\epsilon$. We have the same result even for negative $\epsilon$
because the Cotton potential $V_{IR}^{(II)}$ is always  zero at the
origin. This means that we could not make a transition from chaotic
to non-chaotic behavior in the 4D phase space. Explicitly,  as is
shown in Fig. 4,  the shapes of potential  near the origin $(0,0)$
is not changed significantly as the parameter $\omega$ is changed
from 100 to 0.01($\epsilon$ from 100 to 1). For $\omega=100,1$
cases, there are no essential differences when  comparing  with  the
IR case of $\omega=\infty$ (Einstein gravity: EG). It is found that
the origin of (0,0) always remains global minimum, regardless of any
value of $\omega$ which regulates the UV effects. The only
difference is the appearance of local maxima $V_{lm}$  as $\omega$
decreases.

\begin{figure}[b]
\includegraphics{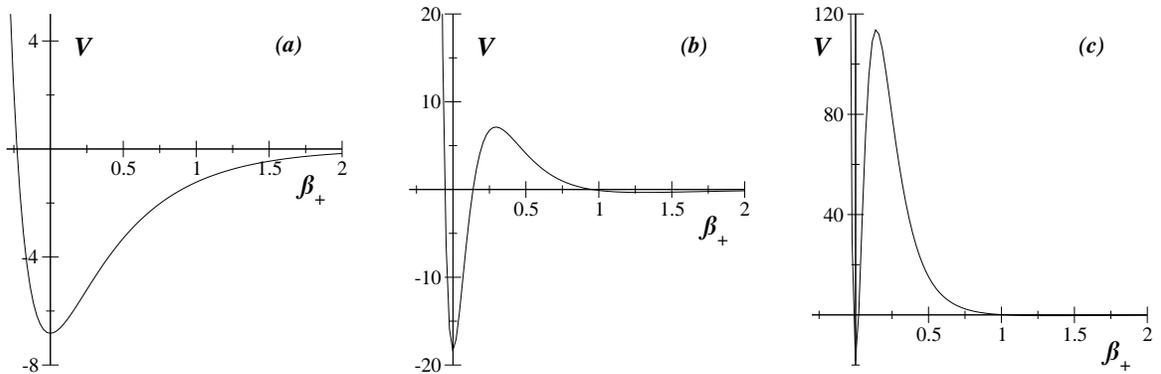}
\caption{Three types of potential graphs
$V(\beta_+,0,\omega,\epsilon)$ for (a) the $\omega=100$ and $
\epsilon=100$ case (EG) without local maximum; (b) the $\omega=0.01 $ and $
\epsilon=100$ case ($z=2$ HL) with local maximum $V_{lm}=7.116$; (c)
the $\omega=0.01$ and $\epsilon=1$ case ($z=3$ HL) with local maximum
$V_{lm}=113.744$. } \label{fig4.eps}
\end{figure}

The chaos could be defined as being such that (i) the periodic
points of the flow associated to the Hamiltonian are dense, (ii)
there is a transitive orbit in the dynamical system, and (iii) there
is sensitive dependence on the initial conditions. Our reduced system
is described by the 4D Hamiltonian
\begin{equation}
  {\cal H}_{4D} =\frac{1}{2}(p^2_++p^2_-) + V(\beta_+,\beta_-,\omega,\epsilon).
\end{equation}
\begin{figure}[t]
\includegraphics{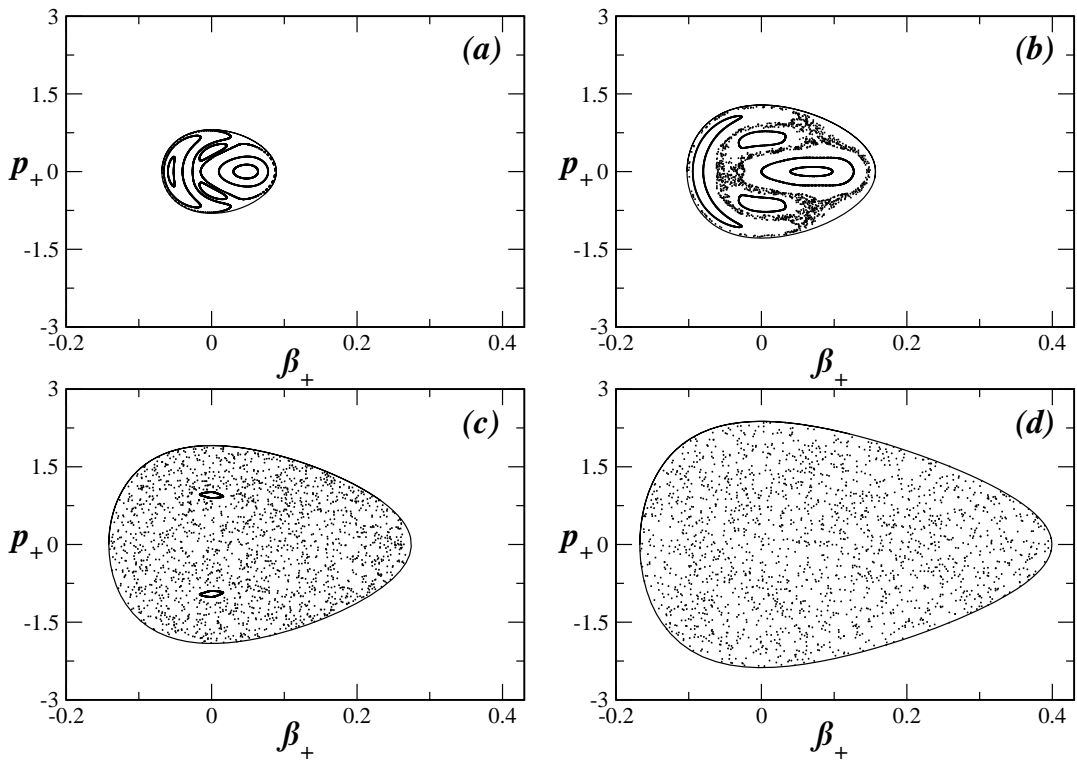}
\caption{Poincar\'{e} sections for the $\omega=100$ and
$\epsilon=100$ case (EG) with (a) $E=-6.5$ (b) $E=-6.0$ (c) $E=-5.0$
(d) $E=-4.0$ } \label{fig5.eps}
\end{figure}

\begin{figure}
\includegraphics{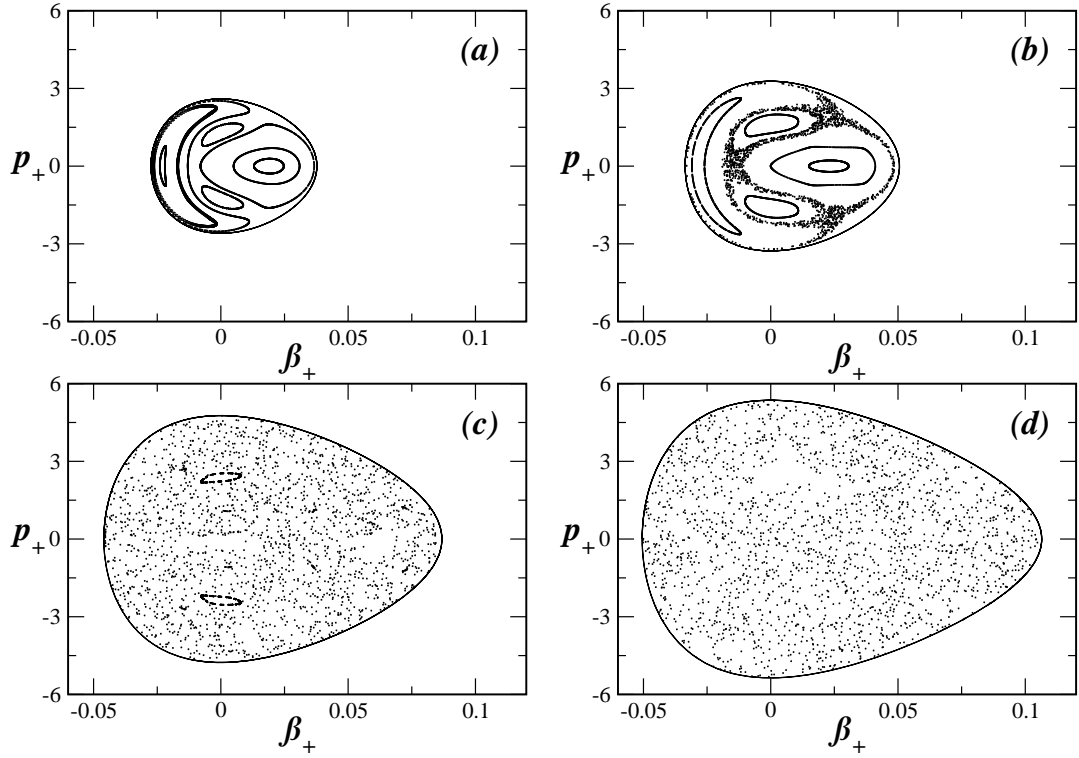}
\caption{Poincar\'{e} sections for the $\omega=0.01$ and $
\epsilon=100$ case ($z=2$ HL gravity) with (a) $ E=-15.0 $(b)
$E=-13.0 $ (c) $E=-7.0$ (d) $E=-4.0$ .} \label{fig6.eps}
\end{figure}

\begin{figure}
\includegraphics{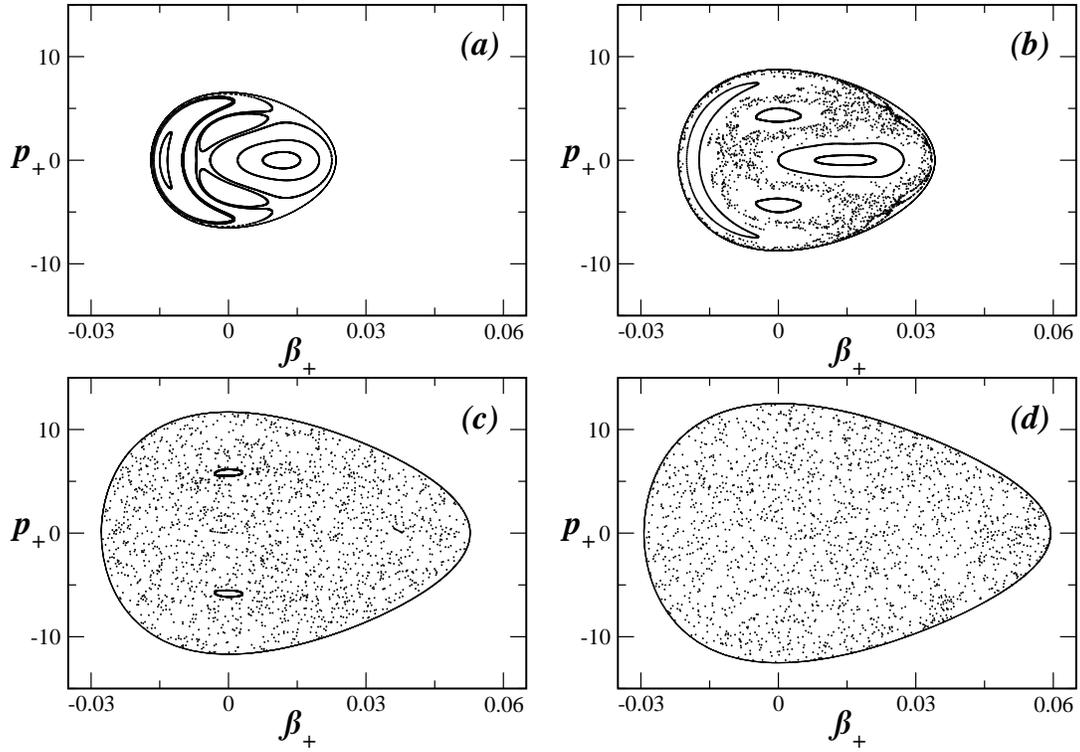}
\caption{Poincar\'{e} sections for the $\omega=0.01$ and $
\epsilon=1$ case ($z=3$ HL gravity) (a) $E=3.0$ (b) $E=20.0$ (c)
$E=50.0$ (d) $E=60.0$. } \label{fig7.eps}
\end{figure}

Now, let us perform simulations of the dynamics and represent
Poincar\'{e} sections, which describe the trajectories in phase
space $(p_+,\beta_+)$ by varying the total energy $E$ or ${\cal
H}_{4D}$ of the system.  We perform the analysis for three
cases: $\omega=100~\epsilon=100$
(EG),~$\omega=0.01~\epsilon=100$($z=2$ HL gravity), and
$\omega=0.01~\epsilon=1$ ($z=3$ HL gravity). We have found that the
chaotic behavior persists for all $\omega>0$. Figs. 5, 6, and  7
represent as Einstein gravity (EG),  $z=2$ HL gravity, and $z=3$ HL
gravity, respectively. These show  that the intersections of several
computed trajectories are displaced in ($p_+,\beta_+$) with
$\beta_-=0$ for different values of energies.  In each plot, we
choose an initial point which corresponds to a prescribed kinetic
energy. The results of Poincar\'{e} sections show that for lower
energy within the potential well, the integrable behavior dominates
and the intersections of trajectories represent closed curves. On
the other hand, for higher energy within the potential well, the
closed curves are broken up gradually and the bounded phase space
fills with a chaotic sea. The same kinds of plots have been obtained
for the other phase space ($p_-,\beta_-$) with $\beta_+=0$.

As a result, we have  obtained that for $\omega>0$, there always
exists chaotic behavior.  This may contrast  to the case of the loop
mixmaster dynamics based on loop quantum cosmology~\cite{Bo}, where
the mixmaster chaos is suppressed by loop quantum effects~\cite{BD}.

However, we have to distinguish the  parameter ``$\omega$" with the
quantum number ``$j$". The former describes the regulation of  UV
effects without spacetime quantization, while the latter depicts the
spacetime quantization and, handles the size of the universe. Hence,
the role of UV coupling parameter $\omega$ is different from the
quantum number $j$ of the loop quantum cosmology. In our case, time
variable (related to the volume of $V=e^{3\alpha}$) as well as two
physical degrees of anisotropy $\beta_{\pm}$ are treated in the
standard way without quantization. However, in the loop quantum
framework, all three scale factors were quantized using the loop
techniques.  Hence two are quite different: the  potential wells at
the origin never disappear for any $\omega>0$ in the $z=3$
Ho\v{r}ava-Lifshitz gravity, while in the loop quantum cosmology the
height of potential wall rapidly decreases until they disappears
completely as the Planck scale is reached. In order to see the
similar effect like decreasing $j$, we consider the volume change
$\alpha$ as the dynamical variable seriously and thus, need a
further work in the 6D phase space.

\section{Chaotic behavior in 6D phase space}
We remind the reader that the true phase space is 6D for the vacuum
universe, and thus, we have a movable billiard with the potential
$V_\alpha(\beta_+,\beta_-,\omega,\epsilon)$ in Eq. (\ref{potalpha})
because the walls are moving with time  since the logarithm of the
volume change $\alpha=\frac{1}{3}\ln V$ and its derivative are
entering in the system.  In this case, $\alpha$ and $p_\alpha$ are
regular variables as functions of time. In this section, we
investigate  a possibility of finding non-chaotic behaviors by
considering  the small volume limit of $\alpha \to -\infty$. To this
end, it would be better to introduce a new time $\tau$ defined
by~\cite{mis}
\begin{equation}
\label{newtime} \tau=\int \frac{dt}{V},~~V=e^{3\alpha},
\end{equation} which makes decoupling of the
volume $\alpha$ from the shape $\beta_\pm$  explicitly. Starting
from the action (\ref{action}) and integrating out the space
variables, we have
\begin{equation}
\label{actionB}
  \bar{S}_{\lambda=1} = (4\pi)^2 \mu^4 \int d\tau
                  \frac{e^{3\alpha}N}{V}
                 \left[6(-\alpha^{'2}+\beta^{'2}_++\beta^{'2}_-)
                    -V^2\left(12e^{-2\alpha}V_{IR}(\beta_+,\beta_-)
                    + \frac{e^{-4\alpha}}{16\omega}V^{(I)}_{UV}(\beta_+,\beta_-)
                    + e^{-6\alpha} V^{(II)}_{UV}(\beta_+,\beta_-)\right) \right],
\end{equation}
where the prime ($'$) denotes the derivatives with respect to
$\tau$. Plugging $N=1$ into (\ref{actionB}), we have the
Lagrangian as
\begin{equation}
\label{Lagtau}
  \bar{\cal L}_{\lambda=1} = (4\pi)^2 \mu^4 \left[6(-\alpha^{'2}+\beta^{'2}_++\beta^{'2}_-)
                    -12e^{4\alpha}\left(V_{IR}(\beta_+,\beta_-)
                    + \frac{e^{-2\alpha}}{192\omega}V^{(I)}_{UV}(\beta_+,\beta_-)
                    + \frac{e^{-4\alpha}}{12} V^{(II)}_{UV}(\beta_+,\beta_-)\right)
                    \right].
\end{equation}
The canonical momenta are given  by
\begin{equation}
  \bar{p}_\pm=\frac{\partial \bar{\cal L}_{\lambda=1}}{\partial\beta'_\pm}
            =12(4\pi)^2\mu^4\beta'_\pm,
  ~~~\bar{p}_\alpha=\frac{\partial \bar{\cal L}_{\lambda=1}}{\partial\alpha'}
            =-12(4\pi)^2\mu^4\alpha'.
\end{equation}
Then, the canonical Hamiltonian in 6D phase space is obtained to be
\begin{eqnarray}
  \bar{\cal H}_{6D}
    &=& \bar{p}_\alpha\alpha'+\bar{p}_+\beta'_++\bar{p}_-\beta'_--\bar{\cal L}_{\lambda=1}\nonumber\\
        &=&   \frac{1}{2}(\bar{p}^2_++\bar{p}^2_--\bar{p}^2_\alpha)
             +e^{4\alpha}\Big(V_{IR}+\frac{e^{-2\alpha}}{192 \omega}V^{(I)}_{UV}
             +\frac{e^{-4\alpha}}{12}V^{(II)}_{UV}\Big),
\end{eqnarray}
where we have chosen the parameter $12(4\pi)^2\mu^4=1$ for
simplicity.  Then, the Hamiltonian  equations of motion are obtained
as
\begin{eqnarray}
 &&\label{appdix1}
 \beta'_\pm=\bar{p}_\pm,
 ~~\bar{p}'_\pm=-e^{4\alpha}\frac{\partial V_{IR}}{\partial\beta_\pm}
               -\frac{e^{2\alpha}}{192\omega}\frac{\partial V^{(I)}_{UV}}{\partial\beta_\pm}
               -\frac{1}{12}\frac{\partial V^{(II)}_{UV}}{\partial\beta_\pm},\\
 &&\label{appdix2}
 \alpha'=-\bar{p}_\alpha,
 ~~\bar{p}'_\alpha =-4e^{4\alpha}V_{IR}
  -\frac{e^{2\alpha}}{96\omega}V^{(I)}_{UV}-\frac{1}{12}V^{CR}_{UV}.
\end{eqnarray}
 We note that comparing Eqs. (\ref{appdix1}) and (\ref{appdix2}) with Eqs. (\ref{betapmeq}) and (\ref{alphaeq}),
  there is no change in the Hamiltonian and its
equations of motion except replacing $t$ by $\tau$.  From Eq.
(\ref{appdix2}), the evolution of $\alpha$ is determined by
\begin{equation}
\alpha''=4e^{4\alpha}V_{IR}
                   +\frac{e^{2\alpha}}{96\omega}V^{(I)}_{UV}
                   +\frac{1}{12}V^{CR}_{UV}.
\end{equation}
Then, we obtain a 6D phase space consisting in the product of a 4D
chaotic one times a 2D regular phase space for the $\alpha$ and
$p_\alpha$ variables.  As the volume goes to zero near  singularity
($e^{4\alpha}\to 0,~p_\alpha \to 0$),  one finds the limit
\begin{equation}
\bar{{\cal H}}_{6D} \to
\frac{1}{2}\Big(\bar{p}_+^2+\bar{p}_-^2\Big)+K \not={\cal H}_{4D}.
\end{equation}
Hence, we note that the 6D system is not asymptotic in  $\tau$ to
the previous 4D system.

Now, we are in a position to  show whether the
presence of  the UV-potential can suppress chaotic behaviors
existing in the IR-potential.  In order to carry out it, we have to
introduce two velocities:  particle velocity $v_p$ and wall velocity
$v_w$ defined by
\begin{equation}
v_p=\sqrt{\bar{p}_+^2+\bar{p}_-^2},~~v_w=\frac{d\beta_+^w}{d\tau},
\end{equation} where the wall location $\beta_+^w$ is determined by
the fact that the asymptotic potential $K$ is significantly felt by
the particle as
\begin{equation}
\label{asymK}
 \bar{p}^2_\alpha\approx 2K=
 \Bigg[\frac{e^{4\alpha-8\beta_+}}{12}+\frac{7e^{2\alpha-16\beta_+}}{32\omega}
 +\frac{4\sqrt{2}e^{\alpha-20\beta_+}}{3\omega^{7/6}\epsilon}+\frac{4e^{-24\beta_+}}{\omega^{4/3}\epsilon^2}
\Bigg]
\end{equation}
in the limit of $\beta_+ \to -\infty$. On the other hand, the
particle velocity is given by
\begin{equation} \label{particlev}
v_p=\sqrt{2\bar{{\cal H}}_{6D}+\bar{p}_\alpha^2-2K}.
\end{equation}
We would like to mention three limiting cases: IR-limit dominated by
$V_{IR}$ and two UV-limits dominated by $V^{CR}_{UV}$ and
$V^{CC}_{UV}$, respectively. In the IR-limit ($\omega \to \infty$)
of Einstein gravity, the wall location is determined by
\begin{equation}
 \beta_+^w \approx \frac{\alpha}{2}-\frac{1}{8}\ln\Big[12\bar{p}_\alpha^2\Big].
\end{equation}
Then, the wall velocity is given by
\begin{equation}
 v_w^{IR}=-\frac{d\beta_+^w}{d\tau} \approx  \frac{\bar{p}_\alpha}{2}+\frac{e^{4\alpha-8\beta_+}}{24\bar{p}_\alpha},
\end{equation}
which leads to
\begin{equation}
|v_w^{IR}| \approx \frac{|\bar{p}_\alpha|}{2}.
\end{equation}
As a result, we find that the particle velocity is always greater than the wall
velocity as
\begin{equation}
v_p^{IR}=\sqrt{2{\bar{{\cal
H}}_{6D}+\bar{p}_\alpha^2-2e^{4\alpha}V_{IR}}} \approx
|\bar{p}_{\alpha}|>v_w^{IR}.
\end{equation}
Thus, there will be an infinite number of collisions of the particle
against the wall since it will always catch a wall~\cite{mix6,mix7}.

Next, let us investigate what happens in the UV-limit.   We mention
that the Cotton bilinear term $C_{ij}C^{ij}$ is marginal in the
$z=3$ Ho\v{r}ava-Lifshitz action and it is expected to dominate in
the UV regime. As its potential $V^{CC}_{UV}$ is shown, it is
independent of volume change $\alpha$. Hence, in this UV regime, one
may approximate Eq. (\ref{appdix2}) to be
\begin{equation}
\alpha'=-\bar{p}_\alpha,~~\bar{p}_\alpha' \approx 0
~~(\alpha''\approx 0),
\end{equation}
which imply that  the scale factor ($V=a_1a_2a_3=e^{3\alpha}$) of
the universe will evolve as a free particle with the fixed momentum
$\bar{p}_\alpha$ and thus, the volume of space diminishes linearly
at early time. Concerning the shape $\beta_\pm$ of the universe,
however,  the potential $V^{CC}_{UV}$ plays no  role in determining
the wall and particle velocities definitely. The wall velocity is
zero as  \be
v^{CC}_w=\frac{d\beta^w_+}{d\tau}=-\frac{1}{12}\frac{\bar{p}'_\alpha}{\bar{p}_\alpha}
\approx 0, \ee while the particle velocity is determined  to be
imaginary
  \be v^{CC}_p \approx
\sqrt{\bar{p}_\alpha^2-\frac{4e^{-24\beta_+}}{\omega^{4/3}\epsilon^2}}
\end{equation}
  for
$\bar{p}_\alpha^2<\frac{4e^{-24\beta_+}}{\omega^{4/3}\epsilon^2}$ in
the limit of $\beta_+ \to -\infty$. In this case, the role of Cotton
bilinear term is  trivial in the 6D phase space.

Finally, we consider the $V^{CR}_{UV}$ term. The wall velocity takes the form
\be |v^{CR}_w|= \frac{|\bar{p}_\alpha|}{20},
\ee
and the particle velocity leads to
\be v^{CR}_p \approx
\sqrt{\bar{p}_\alpha^2-\frac{4\sqrt{2}e^{\alpha-20\beta_+}}{3\omega^{7/6}\epsilon}}\approx
|\bar{p}_\alpha|>|v_w^{CR}|
\ee
in the limit of $\alpha \to -\infty$.
This case is similar to the IR-limit of Einstein gravity.

In summary, we could not observe a slowing down of the particle
velocity due to the UV effects. However, similar to the Einstein gravity,
the mixmaster universe of the $z=3$ deformed Ho\v{r}ava-Lifshitz gravity
filled with stiff matter ($w=1$) has led to a non-chaotic universe
because there is a slowing down of particle velocity  which is
unable to reach any more the walls after some time in the moving
wall picture~\cite{mix6}.

\section{Discussions}

First of all, we wish to mention that  the mixmaster universe has
provided another example that the Ho\v{r}ava-Lifshitz gravity has
shown chaotic behavior, as other chaotic dynamics of string or
M-theory cosmology models~\cite{DH}. This may be because we did not
quantize the Ho\v{r}ava-Lifshitz gravity and we have studied  its
classical aspects only.

The two relevant parameters, which characterize the $z=3$
Ho\v{r}ava-Lifshitz gravity, are are $\omega$ and $\epsilon$. In the
reduced 4D phase space (static billiard), there is no essential
difference in the potentials between $z=1$ (Einstein gravity) and
$z=3$ Ho\v{r}ava-Lifshitz gravity except the appearance of local
maxima. Unfortunately, the local maxima does not change the chaotic
motion significantly and thus, the chaotic behaviors persist in the
$z=3$ Ho\v{r}ava-Lifshitz gravity without a matter.   In the 6D
phase space (movable billiard), the important issue  was  to see
whether the potential $V^{CC}_{UV}$  from the Cotton bilinear term
could slow down the particle velocity $v_p$ relative to the wall
velocity $v_w$.  However, we could not observe a slowing down of the
particle velocity.

At this stage, we compare our results with the mixmaster universe in
the generalized uncertainty principle (GUP)~\cite{BM}. Considering a
close connection between the $z=2$ Ho\v{r}ava-Lifshitz gravity and
GUP~\cite{Myungch}, there may exist a cosmological relation between
them. The chaotic behavior of the Bianchi IX model, which was not
tamed by GUP effects, means that the deformed mixmaster universe is
still a chaotic system. This is mainly because two physical degrees
of anisotropy $\beta_{\pm}$ are considered as deformed, while the
time variable is treated in the standard way. This supports that our
approach (without quantization) is correct.

Furthermore, it was shown that  adding $(^{4}R)^2$ (and possibly
other) curvature terms to the general relativity leads to the
interesting result that the chaotic behavior is
absent~\cite{BC1,BC2,BC3}. Hence it is very curious to see why
$(^{4}R)^2$ does suppress chaotic behavior but
$\frac{3}{4\omega}R^2-\frac{2}{\omega}R_{ij}R^{ij}$ does not
suppress chaotic behavior. In the latter case, $f(R)$-action may
be appropriate for this purpose~\cite{kluson}.

Consequently,  the presence of  the UV-potentials from the $z=3$
deformed Ho\v{r}ava-Lifshitz gravity cannot suppress chaotic
behaviors existing in the IR-potential, which comes from the
Einstein gravity.

\begin{acknowledgments}
Y. S. Myung was supported by Basic Science Research Program
through the National Research Foundation (NRF) of Korea funded by
the Ministry of Education, Science and Technology (2009-0086861).
Y.-W. Kim was supported by the Korea Research Foundation Grant
funded by Korea Government (MOEHRD): KRF-2007-359-C00007. W.-S.
Son and Y.-J. Park were  supported by the Korea Science and
Engineering Foundation (KOSEF) grant funded by the Korea
government (MEST) through WCU Program (No. R31-20002).
\end{acknowledgments}


\begin{thebibliography}{99}

\bibitem{ho1}
  P.~Horava,
  {\it Membranes at Quantum Criticality},
  JHEP {\bf 0903} (2009)  020
  [arXiv:0812.4287 [hep-th]].

\bibitem{ho2}
P.~Horava,
  {\it Quantum Gravity at a Lifshitz Point},
  Phys.\ Rev.\  D {\bf 79} (2009) 084008
  [arXiv:0901.3775 [hep-th]].

\bibitem{ho3}
  P.~Horava,
  {\it Spectral Dimension of the Universe in Quantum Gravity at a Lifshitz Point},
  Phys.\ Rev.\ Lett.\  {\bf 102} (2009) 161301
  [arXiv:0902.3657 [hep-th]].

  \bibitem{CNPS}
  C.~Charmousis, G.~Niz, A.~Padilla and P.~M.~Saffin,
  {\it Strong coupling in Horava gravity},
  JHEP {\bf 0908} (2009) 070
  [arXiv:0905.2579 [hep-th]].

\bibitem{LP}
  M.~Li and Y.~Pang,
  {\it A Trouble with Ho\v{r}ava-Lifshitz Gravity},
  JHEP {\bf 0908} (2009) 015
  [arXiv:0905.2751 [hep-th]].

  \bibitem{SVW}
  T.~P.~Sotiriou, M.~Visser and S.~Weinfurtner,
  {\it Quantum gravity without Lorentz invariance},
  JHEP {\bf 0910} (2009) 033
  [arXiv:0905.2798 [hep-th]].



\bibitem{BPS1}
  D.~Blas, O.~Pujolas and S.~Sibiryakov,
  {\it On the Extra Mode and Inconsistency of Horava Gravity},
  JHEP {\bf 0910} (2009) 029
  [arXiv:0906.3046 [hep-th]].

\bibitem{KA}
  K.~Koyama and F.~Arroja,
  {\it Pathological behaviour of the scalar graviton in Ho\v{r}ava-Lifshitz gravity},
  arXiv:0910.1998 [hep-th].

\bibitem{HKG}
  M.~Henneaux, A.~Kleinschmidt and G.~L.~Gomez,
  {\it A dynamical inconsistency of Horava gravity},
  arXiv:0912.0399 [hep-th].

  \bibitem{M-massive}
  Y.~S.~Myung,
  {\it Propagations of massive graviton in the deformed Ho\v{r}ava-Lifshitz gravity},
  arXiv:0906.0848 [hep-th], to appear in Physical review D.

\bibitem{BPS2}
  D.~Blas, O.~Pujolas and S.~Sibiryakov,
  {\it A healthy extension of Horava gravity},
  arXiv:0909.3525 [hep-th].

\bibitem{PS}
  A.~Papazoglou and T.~P.~Sotiriou,
  {\it Strong coupling in extended Horava-Lifshitz gravity},
  Phys.\ Lett.\  B {\bf 685} (2010) 197
  [arXiv:0911.1299 [hep-th]].

\bibitem{BPS3}
  D.~Blas, O.~Pujolas and S.~Sibiryakov,
  {\it Comment on `Strong coupling in extended Horava-Lifshitz gravity'},
  arXiv:0912.0550 [hep-th].



\bibitem{cal}
   G.~Calcagni,
  {\it Cosmology of the Lifshitz universe},
  JHEP {\bf 0909} (2009)  112
  [arXiv:0904.0829 [hep-th]].


  \bibitem{KK}
  E.~Kiritsis and G.~Kofinas,
  {\it Horava-Lifshitz Cosmology},
  Nucl.\ Phys.\  B {\bf 821} (2009)  467
  [arXiv:0904.1334 [hep-th]].

  \bibitem{muk}
  S.~Mukohyama,
  {\it Scale-invariant cosmological perturbations from Horava-Lifshitz gravity without inflation},
  JCAP {\bf 0906} (2009)  001
  [arXiv:0904.2190 [hep-th]].


\bibitem{TS}
  T. Takahashi and J. Soda,
  {\it Chiral primordial gravitational waves from a Lifshitz point},
  Phys. Rev. Lett. {\bf 102} (2009)  231301
  [arXiv:0904.0554 [hep-th]].


\bibitem{Bra}
  R.~Brandenberger,
  {\it Matter Bounce in Horava-Lifshitz Cosmology},
  Phys. Rev. D{\bf 80} (2009)  043516
  [arXiv:0904.2835 [hep-th]].


  \bibitem{Rama}
  S.~Kalyana Rama,
  {\it Anisotropic Cosmology and (Super)Stiff Matter in Ho\v{r}ava's Gravity Theory},
  Phys.\ Rev.\  D {\bf 79} (2009)  124031
  [arXiv:0905.0700 [hep-th]].

\bibitem{LS}
  G.~Leon and E.~N.~Saridakis,
  {\it Phase-space analysis of Horava-Lifshitz cosmology},
  JCAP {\bf 0911} (2009)  006
  [arXiv:0909.3571 [hep-th]].



\bibitem{KS}
  A.~Kehagias and K.~Sfetsos,
  {\it The black hole and FRW geometries of non-relativistic gravity},
  Phys.\ Lett.\  B {\bf 678} (2009)  123
  [arXiv:0905.0477 [hep-th]].

\bibitem{Myungbh}
  Y.~S.~Myung,
  {\it Thermodynamics of black holes in the deformed Ho\v{r}ava-Lifshitz gravity},''
  Phys.\ Lett.\  B {\bf 678} (2009)  127
  [arXiv:0905.0957 [hep-th]].


\bibitem{mix1} C.~W.~Misner,
  {\it Mixmaster universe},
  Phys.\ Rev.\ Lett.\  {\bf 22} (1969)  1071.

\bibitem{mix2} B.~L.~Hu and T.~Regge,
  {\it Perturbations on the Mixmaster Universe},
  Phys.\ Rev.\ Lett.\  {\bf 29} (1972)  1616.

\bibitem{mix3} D.~F.~Chernoff and J.~D.~Barrow,
  {\it Chaos In The Mixmaster Universe},
  Phys.\ Rev.\ Lett.\  {\bf 50} (1983)  134.

\bibitem{mix4} N.~J.~Cornish and J.~J.~Levin,
  {\it The mixmaster universe is chaotic},
  Phys.\ Rev.\ Lett.\  {\bf 78} (1997)  998
  [arXiv:gr-qc/9605029].

\bibitem{cl} N.~J.~Cornish and J.~J.~Levin,
  {\it The mixmaster universe: A chaotic Farey tale},
  Phys.\ Rev.\  D {\bf 55} (1997) 7489
  [arXiv:gr-qc/9612066].


\bibitem{mix5} M.~P.~Dabrowski,
  {\it Kasner asymptotics of mixmaster Horava-Witten cosmology},
  Phys.\ Lett.\  B {\bf 474} (2000)  52
  [Phys.\ Lett.\  B {\bf 496} (2000)  226]
  [arXiv:hep-th/9911217].

\bibitem{mix6}   T.~Lehner and L.~Di Menza,
  {\it Revisitation of chaos in Bianchi IX universe and in generalized scalar-tensor cosmologies},
  Chaos Solitons Fractals {\bf 16} (2003)  597.


\bibitem{mix7} L.~Di Menza and T.~Lehner,
  {\it The Chaotic Mixmaster And The Suppression Of Chaos In Scalar-Tensor Cosmologies},
  Gen.\ Rel.\ Grav.\  {\bf 36} (2004)  2635.

\bibitem{BKL} V.~A.~Belinsky, I.~M.~Khalatnikov and E.~M.~Lifshitz,
  {\it Oscillatory approach to a singular point in the relativistic cosmology},
  Adv.\ Phys.\  {\bf 19} (1970)  525.

  \bibitem{BD}
  M.~Bojowald and G.~Date,
  {\it A non-chaotic quantum Bianchi IX universe and the quantum structure of
  classical singularities},
  Phys.\ Rev.\ Lett.\  {\bf 92} (2004)  071302
  [arXiv:gr-qc/0311003].

\bibitem{Bo}
  M.~Bojowald,
  {\it Loop quantum cosmology},
  Living Rev.\ Rel.\  {\bf 8} (2005)  11
  [arXiv:gr-qc/0601085].

  \bibitem{MKSP}
  Y.~S.~Myung, Y.~W.~Kim, W.~S.~Son and Y.~J.~Park,
  {\it Chaotic universe in the z=2 Hovava-Lifshitz gravity},
  arXiv:0911.2525 [gr-qc].

\bibitem{BBLP}
  I.~Bakas, F.~Bourliot, D.~Lust and M.~Petropoulos,
  {\it Mixmaster universe in Horava-Lifshitz gravity},
  Class.\ Quant.\ Grav.\  {\bf 27} (2010) 045013
  [arXiv:0911.2665 [hep-th]].



\bibitem{adm} R.L. Arnowitt, S. Deser and C.W. Misner,
{\it The dynamics of general relativity,} ``Gravitation: an
introduction to current research'', Louis Witten ed. (Wilew 1962),
chapter 7, pp 227-265, arXiv:gr-qc/0405109.


\bibitem{Myungch}
  Y.~S.~Myung,
  {\it Chiral gravitational waves from z=2 Ho\v{r}ava-Lifshitz gravity},
  Phys.\ Lett.\  B {\bf 684} (2010) 1
  [arXiv:0911.0724 [hep-th]].



\bibitem{dhn}
  T.~Damour, M.~Henneaux and H.~Nicolai,
  {\it Cosmological billiards},
  Class.\ Quant.\ Grav.\  {\bf 20} (2003)  R145
  [arXiv:hep-th/0212256].



\bibitem{DH}
  T.~Damour and M.~Henneaux,
  {\it Chaos in superstring cosmology},
  Phys.\ Rev.\ Lett.\  {\bf 85} (2000)  920
  [arXiv:hep-th/0003139].

\bibitem{BM}
  M.~V.~Battisti and G.~Montani,
  {\it The Mixmaster Universe in a generalized uncertainty principle framework},
  Phys.\ Lett.\  B {\bf 681} (2009)  179
  [arXiv:0808.0831 [gr-qc]].


\bibitem{BC1}
  J.~D.~Barrow and H. Sirousse-Zia,
  {\it Mixmaster cosmological model in theories of gravity with a quadratic lagrangian},
  Phys.\ Rev. \ D {\bf 39} (1989) 2187.

\bibitem{BC2}
  J.~D.~Barrow and S.~Cotsakis,
  {\it Chaotic behavior in higher order gravity theories},
  Phys.\ Lett.\  B {\bf 232} (1989) 172.

\bibitem{BC3}
  S. Cotsakis, J. Demaret, Y. de Rop and L. Querella,
  {\it Mizmaster universe in fourth-order gravity theories},
  Phys.\ Rev. \ D {\bf 48} (1993) 4595.


\bibitem{mis} C.W. Misner, {\it Deterministic Chaos in General Relativity},
"Chaos in the Einstein equations-characterization and importance",
D. Hobill, A. Burd and A. Coley ed. (Plenum, New York 1991), pp. 317-328.


\bibitem{kluson}
  J.~Kluson,
  {\it Horava-Lifshitz f(R) Gravity},
  JHEP {\bf 0911} (2009) 078
  [arXiv:0907.3566 [hep-th]].




\end{thebibliography}
\end{document}